# ReactiFi: Reactive Programming of Wi-Fi Firmware on Mobile Devices

Artur Sterz[a], Matthias Eichholz[b], Ragnar Mogk[b], Lars Baumgärtner[b], Pablo Graubner[a], Matthias Hollick[b], Mira Mezini[b], and Bernd Freisleben[a]

a Department of Mathematics and Computer Science, Philipps-Universität Marburg, Germany

b Department of Computer Science, TU Darmstadt, Germany

**Abstract** Network programmability will be required to handle future increased network traffic and constantly changing application needs. However, there is currently no way of using a high-level, easy to use programming language to program Wi-Fi firmware. This impedes rapid prototyping and deployment of novel network services/applications and hinders continuous performance optimization in Wi-Fi networks, since expert knowledge is required for both the used hardware platforms and the Wi-Fi domain. In this paper, we present ReactiFi, a high-level reactive programming language to program Wi-Fi chips on mobile consumer devices. ReactiFi enables programmers to implement extensions of PHY, MAC, and IP layer mechanisms without requiring expert knowledge of Wi-Fi chips, allowing for novel applications and network protocols. ReactiFi programs are executed directly on the Wi-Fi chip, improving performance and power consumption compared to execution on the main CPU. ReactiFi is conceptually similar to functional reactive languages, but is dedicated to the domain-specific needs of Wi-Fi firmware. First, it handles low-level platform-specific details without interfering with the core functionality of Wi-Fi chips. Second, it supports static reasoning about memory usage of applications, which is important for typically memory-constrained Wi-Fi chips. Third, it limits dynamic changes of dependencies between computations to dynamic branching, in order to enable static reasoning about the order of computations. We evaluate ReactiFi empirically in two real-world case studies. Our results show that throughput, latency, and power consumption are significantly improved when executing applications on the Wi-Fi chip rather than in the operating system kernel or in user space. Moreover, we show that the high-level programming abstractions of ReactiFi have no performance overhead compared to manually written C code.

## 1 Introduction

Software-Defined Networking (SDN) [31] has revolutionized wired networks by making networking hardware programmable. Software-Defined Wireless Networks (SDWN) [14] promise a similar revolution for wireless networks, but the technology typically runs on access points (APs) or middleboxes. There is no support for programming Wi-Fi chips of mobile end user devices. We argue that (i) programmability of Wi-Fi chips supports the development of novel functionality in an efficient manner, and (ii) high-level dataflow languages are needed for this purpose.

Regarding (i), Cisco predicts that traffic from mobile end user devices will account for more than 54 % of the total IP traffic by 2022 and will continue to grow [12]. Handling this traffic directly on the Wi-Fi chip rather than in the operating system – or, even worse, in the user space of a mobile device – aids in optimizing both power consumption and execution performance. Furthermore, programmable Wi-Fi chips create opportunities for new network services and applications; examples include applications that require sophisticated PHY layer channel selection algorithms, novel shared medium MAC layer access protocols, or IP layer routing and topology creation.

Regarding (ii) – support for high-level languages – we split our argument in two related parts. The first part concerns the interface between applications and Wi-Fi chips. The second part concerns the high-level languages we find appropriate for the Wi-Fi domain.

For the first part, we argue that Wi-Fi chips should be programmed at a high level of abstraction without expert knowledge and in a platform independent way. Both Wi-Fi chips and applications for them are quite diverse and have to adapt to quickly changing requirements. More importantly, we have to ensure that applications do not disturb basic functionality that Wi-Fi chips must provide according to their technical specifications. For example, the IEEE 802.11 specification states that Wi-Fi frames must be acknowledged by the receiver within 30 µs to 50 µs. Such requirements are hard to guarantee if programmers are exposed to low-level details of Wi-Fi chips.

For the second part, we argue that Wi-Fi chips should be programmed with dataflow languages because the explicit dataflow is key in enabling the language machinery to take care of domain-specific properties and to perform domain-specific optimizations. Imperative languages are not well-suited for applications that handle network traffic, since these are typically triggered on event occurrences like data arrival. The problem is the mismatch between the event-driven flow of such applications and the control-driven nature of imperative languages. Callbacks, typically used to work around this mismatch, do not only cause designs to be fragile and hard to maintain [17, 35, 37], but they also make it hard to statically reason about the runtime behavior of programs. This is particularly important for resourced-constrained platforms such as Wi-Fi chips.

We support our arguments (i) by presenting a domain-specific language (DSL), called ReactiFi, dedicated to programming Wi-Fi chips on mobile end user devices, and (ii) by using ReactiFi to implement case studies to qualitatively and empirically validate our claims. ReactiFi is conceptually similar to functional reactive programming languages, e. g., FrTime [13] and REScala [36]. Like these languages, it provides programming abstractions for defining computations that are automatically triggered on data arrival and can be composed using functional combinators. Like other reactive languages, the implementation translates a ReactiFi program into a dataflow graph (DG) that explicitly models the dependencies between computations. ReactiFi differs from other reactive languages in features that address specific needs of the Wi-Fi firmware. The ReactiFi compiler handles platform-specific compilation, bindings to platform-specific APIs, and dynamic loading into the Wi-Fi firmware. As a result, it is possible to program Wi-Fi functionality in a platform-independent manner at a high-level of abstraction that disables modifications of core parts of Wi-Fi chips to guarantee their basic operation. ReactiFi offers only fixed-size types, which together with its event-based dataflow programming model enable static reasoning about memory usage of applications. Furthermore, ReactiFi limits dynamic changes of dependencies between computations to dynamic branching – the static DG enables static reasoning about the order of computations.

ReactiFi programs can use PHY, MAC, and IP layer mechanisms, e. g., reception of frames or changed radio link properties, to enable more efficient and novel networking functionalities. We present two case studies to demonstrate the benefits of ReactiFi in realistic applications. The first one is about counting nearby devices, e. g., for contact tracing. The second one implements new functionality – an adaptive file sharing application by dynamically switching to the most suitable Wi-Fi communication mode. We show that (i) by executing these applications on the Wi-Fi chip, power consumption can be reduced by up to 87 %, and (ii) by exploiting information that is only available on the Wi-Fi chip, data throughput is increased by a factor of up to 3.3 in the adaptive file sharing scenario. Our case studies also demonstrate the advantages of using a high-level dataflow language: the ReactiFi program for adaptive file sharing is 9x shorter than a corresponding low-level C program. Moreover, it is platform independent and has a clear design structure with explicit dataflow paths that are easy to follow and reason about. Finally, the benefits of ReactiFi abstractions in terms of code complexity and platform independence come without regrets: our empirical evaluation shows that there is no runtime performance overhead compared to manually written C code.

To summarize, our contributions are:

- the design of the ReactiFi language, a high-level domain-specific reactive language for programs running on Wi-Fi chips (section 2).
- a formal semantics and type system for ReactiFi (section 3),
- a compiler designed for resource-constrained Wi-Fi target platforms, supporting dynamic configuration of Wi-Fi firmware for disruption-free loading of new functionality (section 4),
- a qualitative discussion of ReactiFi's benefits in terms of prevented and mitigated problems compared to the state of the art (section 5),
- an empirical evaluation of ReactiFi's performance in real-world settings (section 6).

In addition, section 7 discusses related work, while section 8 concludes the paper and outlines areas for future work.

## 2 ReactiFi by Example

We introduce ReactiFi's concepts informally by discussing implementations of two case studies that are also used in our evaluations in section 6. The first case study implements functionality that - without a programmable Wi-Fi chip – would run in the operating system kernel or in user space - with significant overhead in terms of power consumption (cf. section 6). The second case study illustrates functionality that relies on information that is not always available in the operating system kernel or in user space.

**Counting Nearby Devices** In this case study, wireless devices in the vicinity of a particular device are counted to estimate the number of people in a certain place. In people gatherings, e. g., sport events or music festivals, device counting provides valuable information for organizers or security staff. Urban planning uses the number of pedestrians and their flows. In emergency cases, knowing how many people are in an affected area can be live-saving [3, 11, 40]. Listing 1 shows the ReactiFi code for the case study. The program counts MAC addresses of Wi-Fi management frames collected in monitor mode on all Wi-Fi channels and sends the number of addresses to the host operating system every 200 ms.

**Listing 1** ReactiFi program for counting nearby devices.

```
1  Source(Timer(10ms))
2    .fold({ 0 })((channel, time) ⇒ { (channel % 20) + 1 })
3    .observe(SwitchChannel)
4  val addrs = Source(Monitor)
5    .filter(frame ⇒ { frame.type == MANAGEMENT })
6    .map(frame ⇒ { frame.src })
7  val timer = Source(Timer(200ms))
8  val count = fold({ hashset_new() })(
9                  timer → (acc, time) ⇒ { hashset_new() },
10                 addrs → (acc, addr) ⇒ { hashset_add(acc, addr) })
11               .map(p ⇒ { sizeof(p) })
12 timer.snapshot(count.change(0))
13   .map(tuple ⇒ { tuple.snd })
14   .observe(SendToOS)
```

A ReactiFi program consists of reactive definitions – called *reactives* – that encode individual processing steps triggered by incoming events. In listing 1, all bold keywords (except *val*) define reactives for operations such as filters, transformations, and aggregations. Reactives may be parameterized with functions, e. g., to specify which values are filtered by a *filter* reactive. Function bodies (in braces) are written using C code embedded in ReactiFi (cf. section 4) – they operate on simple values, may access only their parameters (but not reactive definitions), and influence the dataflow only via return values. Reactives can be given names (*val*) and they can be composed (via the "." notation) into an acyclic dataflow graph (visualized in figure 2).

Lines 1-3 switch through all Wi-Fi channels. We use a time-based event source that triggers an event every 10 ms. The *fold* reactive aggregates state given an initial value, i. e., it counts how often the source has triggered an event to compute the channel that should be selected. This *fold* reactive then propagates the channel number to the *observe* reactive that executes the *SwitchChannel* side effect. *SwitchChannel* instructs the Wi-Fi firmware to switch to the provided Wi-Fi channel. Line 4 shows a reactive *addrs* that is derived from a chain of reactives to gather all Wi-Fi frames in monitor mode, filter out all non-management frames (line 5), and then project the source MAC address field using *map* (line 6). We use a second timer to report the number of distinct addresses seen within the last 200 ms (line 7). The total count is obtained through a *fold* reactive with multiple triggering conditions, one on *timer* and one on *addrs* (lines 8-10). Both parts refer to the same aggregated state, initialized with an empty set (*hashset_new()*). When *timer* triggers, the set is reset to become empty again, and when *addrs* triggers, the source addresses are collected in that set. When both trigger, the functions are executed in their defined order. The *count* (line 12) maps the accumulated set to its size. A *snapshot* reactive reads the current value of *fold* reactives, when another reactive triggers, i. e., in line 12, a snapshot of *count* is taken, when *timer* triggers. However, the *timer* reactive also resets *count* to zero. To report the count before the *timer* reactive triggers, we use *change*. Using *change* produces a reactive that reports both the current and the previous value, such that programmers may reason about what happened before the current time step. The parameter passed to *change* is used as the initial value of *fold*, e. g., zero in our example. In lines 13 and 14, the *map* reactive then propagates the device count to the *observe* reactive that executes the *SendToOS* side effect. *SendToOS* instructs the Wi-Fi firmware to send data back to the operating system.

**Adaptive File Sharing** In this case study, we use ReactiFi to implement file sharing directly on the Wi-Fi chip. File sharing in local wireless networks is a common service, e. g., Apple “AirDrop”, Microsoft “Nearby Sharing”, and Google “Nearby” are APIs and functionalities for such applications. Unlike these solutions, however, our file sharing application automatically switches from a direct connection between Wi-Fi devices (IEEE 802.11z Tunneled Direct Link Setup (TDLS)) to routing over the nearest access point (IEEE 802.11n AP mode), if that connection is better. This improves throughput and reduces Wi-Fi congestion.

Figure 1 depicts our scenario. When a file is distributed in a local wireless network, the sender as the source of the file transmits the data to an AP that relays it to the receiving destination of the file, resulting in two data streams with the same payload. If the receiver detects that the sender is in close proximity (as it is the case in the beginning and at the end of the scenario timeline in figure 1), it switches from AP to TDLS to establish a direct communication tunnel to the Wi-Fi device, without losing or disturbing the previously established connection to the AP. This kind of application has to be implemented on the Wi-Fi chip, since the required information is not accessible in the operating system, and the kernel cannot switch from 802.11n to 802.11z.

Listing 2 shows the ReactiFi code. On a high level, the application computes the signal-to-noise ratio (SNR) (line 15) for other devices and each supported mode (direct or via AP). The SNRs are stored in a hashmap based on a key derived from the source address of the frame and whether it is a direct frame or a routed frame. To compute the keys, we first use *filter* reactives (lines 7 and 10) to separate the frames into the two types, then use *map* reactives to generate keys from each type of frame. The two types of keys are combined using the *choice* reactive (||) in line 12. *Choice* reactives propagate the value whenever either the left or right reactive triggers, using the left input if both trigger. Then, the SNRs for the source address of the current frame are fetched from the hashmap (line 24) and, depending on which mode we are in (*c1* or *c2*), we compute a boolean on whether or not TDLS should be enabled, which is then sent to the ReactiFi runtime to ensure correct execution (line 28).

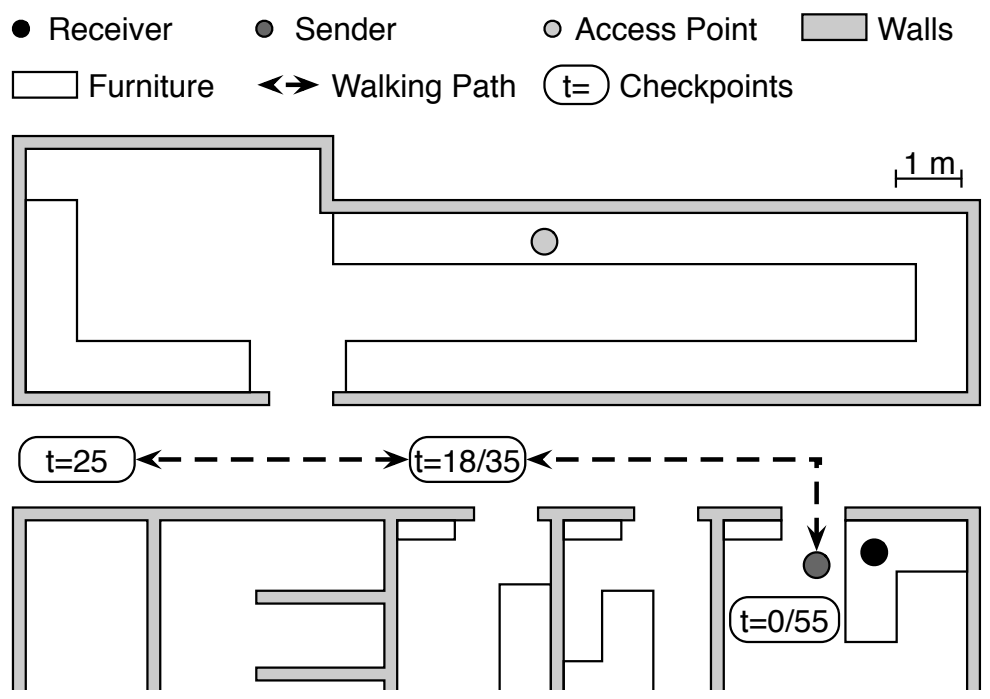

**Figure 1** Scenario for the adaptive file sharing case study. The sender walks back and forth starting (t=0) and ending (t=55) near the receiver.

**Listing 2** ReactiFi program for adaptive file sharing.

```
1  def src_key = frame => { compound_key(frame.src, FROM_SRC) }
2  def ap_key = frame => { compound_key(frame.src, FROM_AP) }
3  val monitor = Source(Monitor)
4  val frames = monitor.filter(frame => { frame.dst == ADDR })
5  val count = frames.fold({ 0 })((count, frame) => { count + 1 })
6
7  val fromSource = frames.filter(frame => {
8    frame.type == FROM_SRC_TO_AP || frame.type == FROM_SRC_TO_DST
9  })
10 val fromAP = frames.filter(frame => { frame.type == FROM_AP_TP_DST })
11
12 val keys = fromSource.map(src_key) || fromAP.map(ap_key)
13 val foreign_keys = fromSource.map(ap_key) || fromAP.map(src_key)
14
15 val avgSnrPerSrc = (count, frames, keys)
16   .fold({ hashmap_new() })((acc, count, frame, key) => {
17     int avg = hashmap_get(acc, key);
18     if (avg == MAP_ENTRY_MISSING) {
19       return hashmap_put(acc, key, frame.snr);
20     } else {
21       return hashmap_put(acc, key, avg + (frame.snr - avg) / count);
22     }
23   })
24 val c1 = (avgSnrPerSrc, fromSource, keys, foreign_keys)
25   .map((avgs, frame, k, fk) => { hashmap_get(avgs, k) > hashmap_get(avgs, fk) })
26 val c2 = (avgSnrPerSrc, fromAP, keys, foreign_keys)
27   .map((avgs, frame, k, fk) => { hashmap_get(avgs, k) < hashmap_get(avgs, fk)})
28 (c1 || c2).observe(SetTDLS)
```

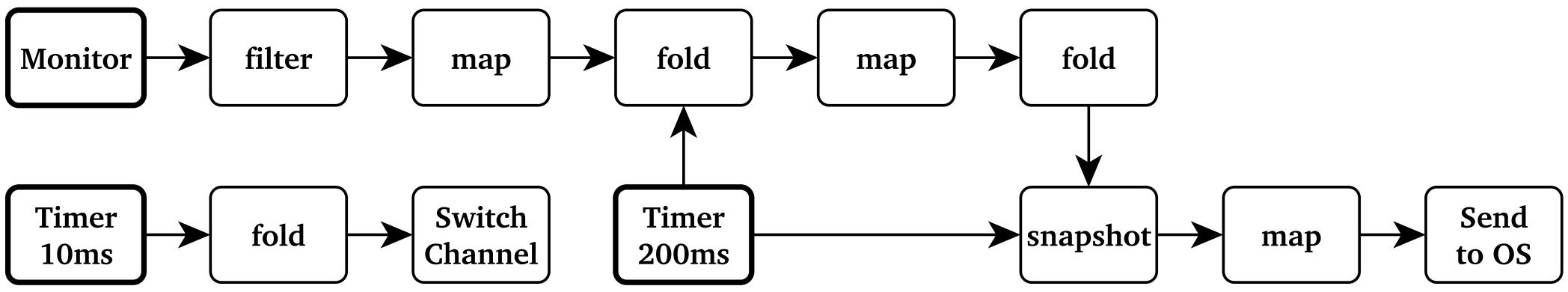

**Figure 2** Dataflow graph of the nearby-device counting case study.

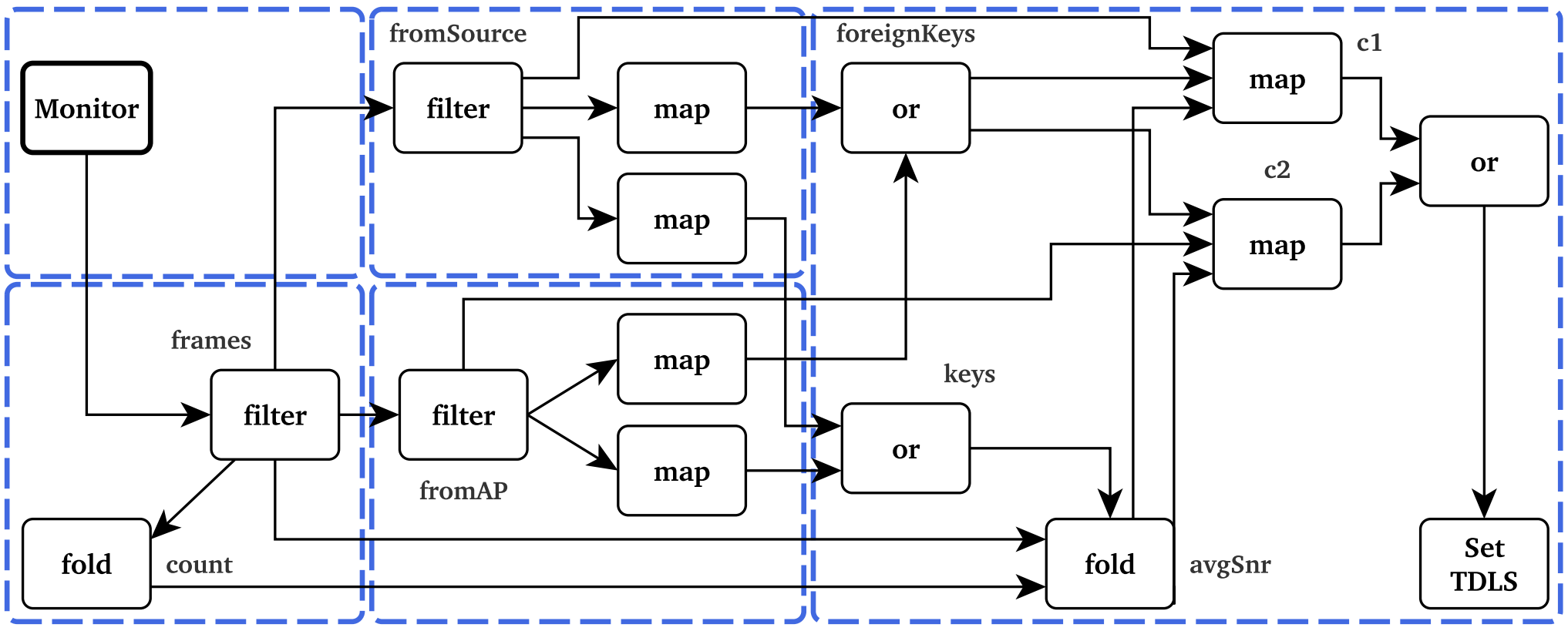

**Figure 3** Dataflow graph of the adaptive file sharing case study.

**Dataflow Graph and Event/Data Propagation** A ReactiFi program is transformed into a dataflow graph (DG) (cf. section 4) that represents the abstract program logic. Each reactive $r$ is a node in the DG with incoming edges from all inputs of $r$. Reactives must be declared before they are used, thus the DG is always acyclic. The DG guides the process of handling incoming events. A source is automatically triggered on arrival of incoming events from the firmware. The reactions are transitively propagated along DG paths, during which derived reactives transform, filter, and aggregate the results of other reactives, or the state in *fold* reactives gets updated. At the end of the propagation process, external effects of triggered observers are executed in the Wi-Fi firmware. For illustration, the DGs of our case studies are shown in figure 2 and figure 3 (the meaning of the blue boxes in figure 3 is explained in section 4).

## 3 The ReactiFi Language: Syntax and Semantics

Figure 4 shows the syntax of ReactiFi. A ReactiFi program is a sequence of definitions $d$ of the kind val $x = r$, each denoting a reactive expression $r$ by an identifier $x$. A reactive $r$ is either a source without inputs, or is derived from its input reactives $\overline{x}$ (shorthand for $(x_1, \ldots, x_n)$) using one of the combinators *map*, *fold*, $||$, *filter*, or *snapshot*. Some combinators are parameterized by an initial value $v$ or a function $f$. Values and function bodies are written in the C programming language. The examples shown in section 2 use syntactic sugar for chained pipelines. The single assignment form used here simplifies the presentation without affecting the semantics.

| | | |
|---|---|---|
| $x$ | ::= *identifiers* | *Variables* |
| $v$ | ::= *fixed size values* | *Values* |
| $f$ | ::= $\overline{x} \Rightarrow \{cblock\}$ | *Functions* |
| $s$ | ::= Monitor \| Timer($v$) \| ... | *Sources* |
| $o$ | ::= SendToOs \| SwitchChannel \| ... | *Effects* |
| $r$ | ::= Source($s$) \| $\overline{x}$.map($f$) \| $\overline{x}$.fold($v$)($f$) \| fold($v$)($\overline{x \to f}$) \| ($x$\|\|$x$) \| $x$.filter($f$) \| $x$.snapshot($x$) \| $\overline{x}$.observe($o$) | *Reactives* |
| $d$ | ::= val $x$ : T = $r$ | *Definitions* |

**Figure 4** Syntax of ReactiFi.

To interface with the Wi-Fi chip, ReactiFi uses a set of predefined interactions. Interactions are wrapped into source reactives ($Source(s)$) or are parameters of observer reactives ($\overline{x}$.observe($o$)). Source interactions include receiving frames, timers, or changed channel state (e. g., if a network connection gets disrupted due to the user's mobility). Observer interactions include transmitting custom Wi-Fi frames, sending packets to the network stack of the host operating system, switching between 2.4 GHz and 5 GHz frequency bands, switching channels, etc. The full list of external interactions is shown in table 1 in the Appendix.

Figure 5 shows the typing rules of ReactiFi reactives. To simplify our presentation, we do not show the typing context for variables. We assume each reactive may access all other reactives defined before, but not after, in the list of definitions, resulting in an acyclic graph of dependencies between reactives. All reactives in ReactiFi have the type Reactive[A] and are parametric over the value they carry, but are never nested. We assign semantics to individual ReactiFi reactives by giving the translation of individual definitions val $x$ : T = $r$ into C-like statements shown in figure 6. Translation is written $C[\![$val $x$ : T = $r]\!]$). Conceptually, the translated statements will be executed in the topological order of the DG, as explained in section 4. In general, each statement first checks if the trigger condition (e. g., $T[\![\overline{x}]\!]$) for its inputs (e. g., $\overline{x}$) are fulfilled, and then updates the current value of the reactive ($x_0$). If any condition for a reactive is false and there is no else branch, then the reactive itself does not trigger, stopping the propagation at this point. Transformation of functions $C[\![f(\overline{x})]\!]$ result in a call to a fresh top-level function definition. Compiling an *identifier* that binds a reactive $C[\![x]\!]$ produces code that accesses the value of that reactive.

A *source reactive* is of type Reactive[A], given that it is triggered by a source $s$ of type SourceDef[A], i. e., the inner type A of the reactive is defined by the inner type of $s$ (rule Source). Similarly, an *observe reactive* has a single input $x$ of type Reactive[A], given that they observe an $o$ of type ObserveDef[A], i. e., the inner type of the input reactive must match the type that is consumed by $o$ (rule Observe). A *filter reactive*

$$\frac{s : \text{SourceDef}[\mathrm{A}]}{\mathsf{Source}(s) : \text{Reactive}[\mathrm{A}]}\ \textsc{Source} \qquad \frac{x : \text{Reactive}[\mathrm{A}] \quad o : \text{ObserverDef}[\mathrm{A}]}{x.\mathsf{observe}(o) : \text{Observer}}\ \textsc{Observe} \qquad \frac{x : \text{Fold}[\mathrm{A}]}{x : \text{Reactive}[\mathrm{A}]}\ \textsc{Subtype}$$

$$\frac{x : \text{Reactive}[\mathrm{A}] \quad f : \mathrm{A} \Rightarrow \text{Boolean}}{x.\mathsf{filter}(f) : \text{Reactive}[\mathrm{A}]}\ \textsc{Filter} \qquad \frac{x_1 : \text{Reactive}[\mathrm{A}] \quad x_2 : \text{Reactive}[\mathrm{A}]}{(x_1 || x_2) : \text{Reactive}[\mathrm{A}]}\ \textsc{Choice}$$

$$\frac{x_i : \text{Reactive}[\mathrm{A_i}] \quad f : \overline{\mathrm{A}} \Rightarrow \mathrm{R}}{\overline{x}.\mathsf{map}(f) : \text{Reactive}[\mathrm{R}]}\ \textsc{Map} \qquad \frac{x_i : \text{Reactive}[\mathrm{A_i}] \quad v : \mathrm{R} \quad f : (\mathrm{R}, \overline{\mathrm{A}}) \Rightarrow \mathrm{R}}{\overline{x}.\mathsf{fold}(v)(f) : \text{Fold}[\mathrm{R}]}\ \textsc{Fold}$$

$$\frac{x_i : \text{Reactive}[\mathrm{A_i}] \quad v : \mathrm{R} \quad f_i : (\mathrm{R}, \mathrm{A_i}) \Rightarrow \mathrm{R}}{\mathsf{fold}(v)(\overline{x \rightarrow f}) : \text{Fold}[\mathrm{R}]}\ \textsc{FoldAll} \qquad \frac{x_1 : \text{Reactive}[\mathrm{A}] \quad x_2 : \text{Fold}[\mathrm{B}]}{x_1.\mathsf{snapshot}(x_2) : \text{Reactive}[\mathrm{B}]}\ \textsc{Snapshot}$$

**Figure 5** Typing rules of reactives in ReactiFi.

$$\begin{aligned}
C[\![\mathsf{val}\ x_0 = \mathsf{Source}(s)]\!] &= \mathsf{if}\ (T[\![s]\!])\{x_0 = C[\![s]\!]\} \\
C[\![\mathsf{val}\ x_0 = \overline{x}.\mathsf{map}(f)]\!] &= \mathsf{if}\ (T[\![\overline{x}]\!])\{x_0 = C[\![f(\overline{x})]\!]\} \\
C[\![\mathsf{val}\ x_0 = \overline{x}.\mathsf{fold}(v)(f)]\!] &= \mathsf{if}\ (T[\![\overline{x}]\!])\{x_0 = C[\![f(x_0, \overline{x})]\!]\} \\
C[\![\mathsf{val}\ x_0 = \mathsf{fold}(v)(x \rightarrow f)]\!] &= \mathsf{if}\ (T[\![x]\!])\{x_0 = C[\![f(x_0, x)]\!]\} \\
C[\![\mathsf{val}\ x_0 = \mathsf{fold}(v)(x_1 \rightarrow f_1, \overline{x \rightarrow f})]\!] &= \mathsf{if}\ (T[\![x_1]\!])\{x_0 = C[\![f_1(x_0, x_1)]\!]\}; \\
&\quad\ C[\![\mathsf{val}\ x_0 = \mathsf{fold}(v)(\overline{x \rightarrow f})]\!] \\
C[\![\mathsf{val}\ x_0 = (x_1 || x_2)]\!] &= \mathsf{if}\ (T[\![x_1]\!])\{x_0 = C[\![x_1]\!]\} \\
&\quad\ \mathsf{else\ if}\ (T[\![x_2]\!])\{x_0 = C[\![x_2]\!]\} \\
C[\![\mathsf{val}\ x_0 = x.\mathsf{filter}(f)]\!] &= \mathsf{if}\ (T[\![x]\!] \&\& C[\![f(x)]\!])\{x_0 = C[\![x]\!]\} \\
C[\![\mathsf{val}\ x_0 = x_1.\mathsf{snapshot}(x_2)]\!] &= \mathsf{if}\ (T[\![x_1]\!])\{x_0 = C[\![x_2]\!]\} \\
C[\![\mathsf{val}\ x_0 = C[\![\overline{x}.\mathsf{observe}(o)]\!]]\!] &= \mathsf{if}\ (T[\![\overline{x}]\!])\{C[\![o(\overline{x})]\!]\}
\end{aligned}$$

**Figure 6** Compiling individual ReactiFi reactives (left) to C code (right).

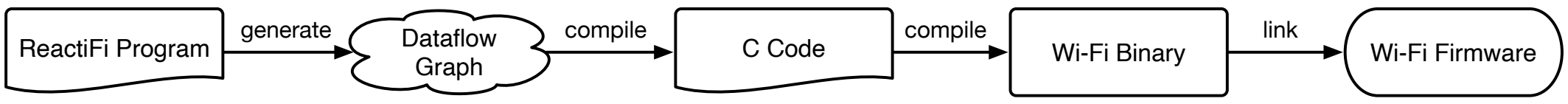

**Figure 7** Stages of compiling ReactiFi to Wi-Fi chips.

has the same type, Reactive[A], as its single input $x$. To use the function $f$ for filtering, the function must take a parameter of the type A and return a Boolean (rule FILTER). *Filter* reactives pass the value of their input unchanged, if the *filter* condition $f(x)$ is true. A *choice reactive* ($||$) takes two inputs $x_1$ and $x_2$ which must be of the same type, and the result is also of that same type (rule CHOICE). It returns the value of its left operand $x_1$ if $x_1$ triggers, otherwise the right operand $x_2$ if $x_2$ triggers. The type of a *map reactive* says that the $i$-th input reactive $x_i$ must match the type of the $i$-th parameter $A_i$ of the given function $f$. The resulting type is Reactive[R] where $R$ is the result type of $f$. For example, a *map* reactive that combines three reactives with types $\mathrm{A}, \mathrm{B}, \mathrm{C}$ as inputs expects a function of type $(\mathrm{A}, \mathrm{B}, \mathrm{C}) \Rightarrow \mathrm{R}$. The value when a map reactive $r$ triggers is function $f$ applied to the value of all inputs $\overline{x}$ of $r$.

A *fold reactive* has type Fold[R] which is a subtype of Reactive[R] (rule SUBTYPE). The Fold[R] type identifies stateful reactives. This type distinction is used to limit which reactives are accessed by *snapshot* reactives. There are two syntactic forms for *fold* reactives. Both are of type Fold[R] with an initial value $v$ of type R. The first syntactic form, $\overline{x}.\mathsf{fold}(v)(f)$, has inputs and a function with matching types, similar to *map* reactives. In addition to *map* reactives, *fold* reactives have their current value $x_0$ stored globally. If all inputs $\overline{x}$ are triggered, then $x_0$ is updated by applying $f$ to $x_0$ and all inputs $\overline{x}$. The second syntactic form, $\mathsf{fold}(v)(\overline{x \to f})$, called *fold all* reactives, has one function per input. *Fold all* reactives are translated to multiple statements, each statements updates the current state $x_0$ by applying $x_0 = f_i(x_0, x_i)$, if the corresponding input $x_i$ is triggered, otherwise the application of that $f_i$ is skipped. All $f_i$ with triggered $x_i$ are applied in the order they are defined, potentially updating $x_0$ multiple times. The typing rule FOLDALL ensures that the second parameter $A_i$ of each $f_i$ matches with the type of the assigned input $x_i$.

A *snapshot reactive* takes two inputs, where the first $x_1$ is a reactive of type Reactive[A] and the second $x_2$ is a *fold* of type Fold[B] and results in a Reactive[B] (rule SNAPSHOT). A *snapshot* reactive triggers when its first input $x_1$ triggers, but returns the value of its second input $x_2$. The type restriction on $x_2$ is because only *fold* reactives have a defined state when they do not trigger.

## 4 The ReactiFi Implementation

A ReactiFi program is processed in four steps (figure 7): (i) the dataflow graph (DG) is constructed, (ii) a C program is generated from the DG, (iii) the C source code is compiled into a binary, which (iv) is loaded to the Wi-Fi chip and linked into the firmware at runtime.

### 4.1 Generating and Typechecking the Dataflow Graph

ReactiFi is implemented as an embedded domain-specific language (EDSL) in Scala, i. e., its abstractions are implemented as a Scala library. We selected Scala due to its support for embedding DSLs, e. g., we implement the DSL to reuse Scala's type checker for the typing rules in figure 5. A ReactiFi program consists of a set of library calls that look like proper DSL syntax. These calls construct the DG that is subsequently compiled to C. User-defined functions are opaque to the Scala DSL – they are directly copied into the generated C code. Once constructed, the DG is analyzed to extract the following information that is passed to the subsequent processing phases: (i) a topological order of all reactives, (ii) a set of conditions guarding the activation of each reactive, and (iii) types and memory requirements of reactives in the DG.

### 4.2 Generating C Code for the Update Function

Given the DG and the C code of the function bodies embedded into reactive definitions, the ReactiFi compiler generates a single sequential update function (UF) in C, which implements the reaction to external events. Thereby, the compiler performs the following domain-specific optimizations:

**Static Sequential Scheduling** The DG specifies a logically concurrent execution order of reactives in response to individual external events; moreover, only a subset of reactives is typically triggered for each external event. Wi-Fi chips only support sequential execution and network applications are often latency-sensitive. To address these constraints, the compiler (a) sequentializes the order of updating reactives and (b) generates a minimum number of conditional branches to select the updated reactives. While the relative execution order of reactives can be statically fixed according to the topological order of them in DG, whether and when reactives trigger depends on runtime conditions. Sources and *filter* reactives define new conditions. For all other reactives, the conditions are derived from the conditions of their input reactives. The *choice* and *fold all* reactives use the disjunction of the conditions of all inputs. All other reactives use the conjunction of the conditions of all inputs. Reactives in the DG are grouped into uninterrupted pipelines based on shared filtering conditions. The compiled update function only checks conditions once per group. For illustration, consider the DG of the file sharing case study in figure 3; the blue boxes mark the uninterrupted groups; for instance, the rightmost group will execute only if the *Monitor* source fires, the condition for the *frames* filter holds, and either *fromSource* or *fromAP* are true.

**Optimized Memory Management** Wi-Fi chips have limited memory. For instance, the memory built into the Nexus 5 used in our evaluation has 768 kB RAM, most of which is used by the basic firmware, with only as little as 100 kB RAM left for higher-level functionality; to put this into context, a single IP packet is up to 2 kB in size. Reactives are abstractions with zero runtime memory cost, i. e., sizeof(Reactive[T]) == sizeof(T). To facilitate compile-time estimation of the needed memory, ReactiFi allows only fixed

**Listing 3** Generated C code example.

```
address_t extractAddress(frame_t fr, frame_t sub) { /* extractAddress */ }
// state of fold reactives would be above
void update() {
  bool monitor_condition;
  frame_t frame_value;
  frame_t subframe_value;
  address_t address_value;

  monitor_condition = runtime_is_triggered(Monitor)
  if (monitor_condition) {
    frame_value = ...;
    subframe_value = ...;
    address_value = extractAddress(frame_value, subframe_value);
    deallocate(frame_value);
    deallocate(subframe_value);
  }
}
```

size types T to be used in code. Memory for reactives is reclaimed at the earliest time possible. Technically, to find the reclamation point of a reactive $r_1$, the compiler traverses the sequential execution order from the back to find the last reactive $r_2$ that depends on $r_1$. The scope of $r_1$ extends from $r_1$ until after $r_2$. Unlike other reactives, the state of *folds* is stored between updates, thus never reclaimed. Overall, for each reactive $r$, the compiler knows how much memory is already allocated when a new value will be computed for $r$. This is the sum of the memory allocated to all *folds* in the program plus the sum of the memory allocated to all non-fold reactives in scope. This way, the compiler is able to maximize the memory available for executing the function bodies embedded in the reactives.

**Exemplary Compilation** For illustration, consider the code below, defining a *map* reactive (*address*) with two inputs, *frame* and *subframe*; we assume that both frames are derived from a *Monitor* source (not shown for brevity).

```
val address = (frame, subframe).map((fr, sub) ⇒ { /* extractAddress */ } )
```

The generated C code is sketched in listing 3. Since there are no folds, the program has only local state. The user-defined function is extracted to a top-level C function `extractAddress`. Within the `update` function, first, the variables for the source conditions (`monitor_condition`), the values computed for the input reactives (`frame_value`, `subframe_value`), and for the *map* reactive (`address_value`) are declared. Then, the trigger conditions of sources are computed, followed by the guarded execution of reactives, when the sources are triggered. The code illustrates the two compiler optimizations. First, the trigger condition is only checked once for all reactives as opposed to once per reactive. Second, the values *frame* and *subframe* are deallocated immediately after they are used and no longer needed. We assume that `address_value` is used later in the program, otherwise the whole program would be optimized away.

### 4.3 From C to Binary to the Wi-Fi Chip

There are several available deployment targets. Some of them, e. g., SoftMAC Wi-Fi dongles [43], or Espressif's ESP platform [32], are special-purpose hardware with custom software. On the contrary, the Nexmon firmware patching framework [41, 42] can also be used and executed on off-the-shelf smartphones. Since we want to deploy ReactiFi programs on off-the-shelf smartphones for validating the feasibility of our approach in real-work case studies, our current implementation uses Nexmon as the target platform. To transfer the compiled ReactiFi program to the Wi-Fi chip, we created a new `ioctl`. Such `ioctl`s are common communication channels between the host (either kernel or user space application) and dedicated hardware components like the Wi-Fi chip. At this point, the dynamic linker usually performs a relocation step to adjust the addresses of branches to their absolute memory location of the loaded code. This relocation step, however, would require the Wi-Fi chip to be restarted, making Wi-Fi communication temporarily unavailable. However, we want to reconfigure the Wi-Fi chip at runtime without disturbing ongoing connections. Therefore, we extended Nexmon to support *Position Independent Code* (PIC). PIC modules can be loaded to arbitrary memory addresses from where their execution can be triggered during runtime. Since the PIC module is unaware of where the binary blob gets loaded, the code performs jumps relative to the program counter. Existing firmware functions, on the other hand, are accessed by first loading the absolute target address from the *Global Offset Table* to a register and then jumping to that address. To recap, our PIC extensions enable loading and executing ReactiFi programs without restarting the Wi-Fi chip. This allows ReactiFi to always ensure basic functionality of the IEEE 802.11 specification.

## 5 Advantages of Using ReactiFi for Wi-Fi Programming Compared to C

In this section, we summarize how ReactiFi's compiler and runtime address domain-specific issues of the Wi-Fi platform. In addition, to help the reader appreciate the benefits of using a high-level dataflow language in terms of code quality, we walk through a C implementation of the adaptive file sharing application shown in the Appendix, and compare the latter with the ReactiFi implementation shown in section 2.

### 5.1 Properties Ensured by the ReactiFi Compiler and Runtime

ReactiFi's declarative dataflow programming model enables the compiler to ensure several properties, as described below.

**Minimized Memory Usage** ReactiFi's programming model matches well the assumption that Wi-Fi chips are supposed to quickly react to incoming events, but only store limited state. First, data pertaining to an event only exists for the duration of the event. Unused or inactive reactives and their derived reactives are not executed or initialized at all and, hence, do not consume any memory. Second, except for *folds*, other reactives do not require to store state between updates. Third, the ReactiFi compiler ensures that temporarily used memory is freed as soon as possible during updates (cf. section 4). Finally, usage of memory by reactives is statically limited in size. ReactiFi only supports parameters to reactive computations with a statically bound size. The usage of any other types is prohibited by the type checker.

**Automatic, Correct, and Optimized Scheduling** The DG allows precise and sound reasoning about the order of reactive computations, enabling compile-time optimizations and scheduling without any runtime overhead. This is possible because ReactiFi limits dynamic changes of the DG and the scheduling order to filtering. It has been argued that the limited expressiveness is sufficient for most programs [54]. As a result, the ReactiFi compiler is free to rearrange the order of execution as long as explicit dependencies between reactives are preserved, allowing to minimize dynamic checks (cf. section 4).

**Platform Independence and Compliance** ReactiFi is compliant with the IEEE 802.11 specification by always providing basic functionality of the Wi-Fi firmware. ReactiFi programs cannot break basic functionality of the Wi-Fi firmware. Interactions only happen through high-level source reactives and observers (cf. section 2 and table 1). Furthermore, the generated code can be deployed on a Wi-Fi chip without interruption (cf. section 4.3). Thus, ReactiFi allows developers with no particular Wi-Fi expertise to write platform-independent Wi-Fi functionality, leaving error-prone and platform-specific aspects to be handled by ReactiFi.

**Non-Ensured Properties** ReactiFi cannot reason about user-defined C code encapsulated in reactives. However, the amount of C code necessary can be kept to a bare minimum, and it is sufficient to review each function individually. While each C function is typechecked by the C compiler, and ReactiFi ensures type correctness of using the function in the DG, ReactiFi cannot ensure that C functions terminate or use a bounded amount of memory. Beyond these type checks and the enforcement of the fixed-sized types, the current type checker of ReactiFi only inherits the standard guarantees of the Scala type checker. In principle, it is possible to extend the type system with user-defined specifications about the behavior of C blocks, e. g., with regard to real-time behavior. With such specifications and the explicit knowledge about the DG, the compiler can reason about real-time guarantees of ReactiFi programs. Such extensions remain to be investigated in future work.

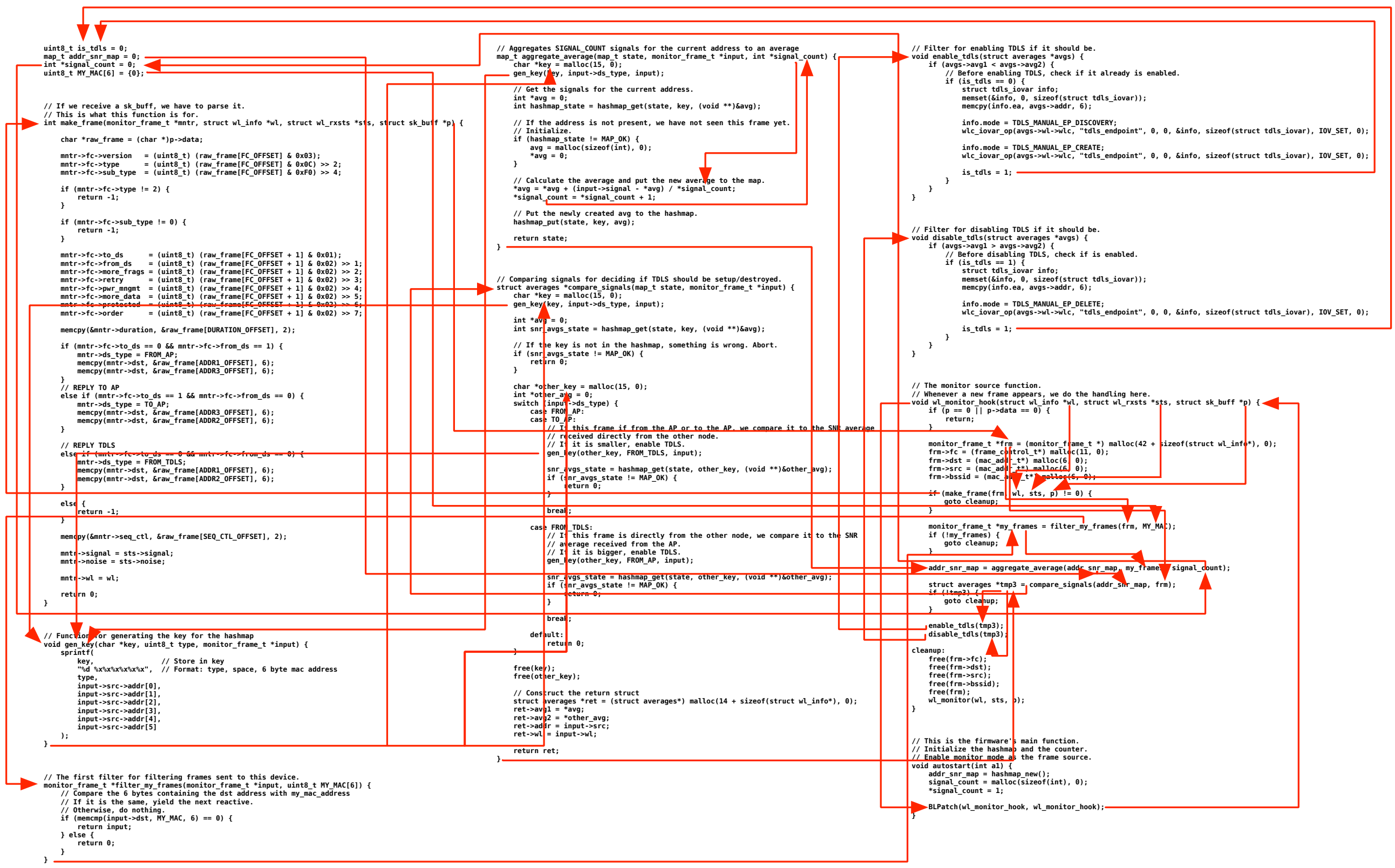

**Figure 8** Dataflow of the C implementation of the adaptive file sharing case study. The arrows represent the direction of the dataflow.

## 5.2 Comparison to C

We compare ReactiFi to C, because C is the most widely used language for programming Wi-Fi chips. Yet, our arguments apply to all imperative languages that do not support a declarative dataflow programming model.[1]

To start with, consider how much code complexity and programming effort ReactiFi saves. While the ReactiFi program is only 28 LoCs long (listing 2), the C program (listing 6 in the Appendix) even without **#include** directives, comments, and empty lines consists of 229 LoCs (almost 9× more!). Moreover, the dataflow of the C implementation shown in figure 8 is over-proportionally complex and hard to reason about with data dependencies modeled implicitly using global state, side effects, and pointers. In contrast, the dataflow of the ReactiFi implementation of the same functionality shown in figure 3 is simple and explicit in the program.

[1] The adaptive file sharing application was implemented in C by one of the authors of the paper, who has ample experience with C programming and the Nexmon platform. While we are aware that this is a threat to the validity of the statements we make in the following, there are no existing programs for the Wi-Fi chip, which we could have used – ultimately, our starting point is that Wi-Fi chips are not programmable today.

Moreover, a programmer that implements Wi-Fi applications in C needs detailed low-level knowledge of the specific platform. Consider as an example the extract from the C implementation of the adaptive file sharing case study in listing 4. To enable monitor mode, the developer must know the memory address of the *Monitor* source for every target Wi-Fi chip and firmware version. Similarly, the developer must often use low-level bitwise operations to implement data extraction from Wi-Fi frames. This requires knowledge of the internal data structures, since information might be stored in different locations inside the frame, depending on the context, e. g., the source and destination addresses of a frame must be extracted differently, depending on whether the frame originates from an access point or not. As an example, listing 5 shows how to get the type field of a Wi-Fi frame.

**Listing 4** Enabling monitor mode using C.

```
__attribute__((at(0x18DA30, "", CHIP_VER_BCM4339, FW_VER_6_37_32_RC23_34_40_r581243)))
__attribute__((at(0x18DB20, "", CHIP_VER_BCM4339, FW_VER_6_37_32_RC23_34_43_r639704)))
BLPatch(wl_monitor_hook, wl_monitor_hook);
```

**Listing 5** Parsing type fields of a Wi-Fi frame in C.

```
mntr->fc->type = (uint8_t) (raw_frame[FC_OFFSET] & 0x0C) >> 2;
mntr->fc->sub_type = (uint8_t) (raw_frame[FC_OFFSET] & 0xF0) >> 4;
```

The above observations indicate that ensuring correctness is difficult in an imperative language like C. It requires highly skilled programmers to understand the program by manually tracking memory allocation along the complex dataflow graph. This makes it very hard for the programmer to assess whether the memory needs of the application can be satisfied by the available memory. Furthermore, incorrect execution order may lead either to memory corruption or inefficient execution. Ensuring that this cannot happen is cumbersome and error-prone: it is necessary to repeatedly test the program to ensure correct order of executions; even providing basic operation requires enormous effort. What is more, the efforts will have to be repeated over and over for any new program, as there is no automated language machinery available. Of course, it is theoretically possible in C to provide higher-level APIs that hide some of the low-level details of the platform. However, such libraries would only partly solve the outlined problems, since providing the memory, scheduling, and platform independence of ReactiFi fundamentally requires to reason about the DG. Furthermore, it is unclear how to ensure that APIs are correctly used [1] and this is especially problematic for APIs abstracting a diverse set of low-level platform-dependent details. At the end, given that ReactiFi's dataflow abstractions come without runtime overhead (as we show in the next section), no good arguments are left for an API-based approach.

## 6 Empirical Evaluation

First, we quantify the basic power consumption and performance of the Wi-Fi chip using a micro-benchmark. Next, we experiment with our two case studies to validate our claims regarding improved power consumption and throughput.

### 6.1 Experiments Using a Micro-benchmark

**The Benchmark** Since currently there are no other programming languages, implementations, or applications for Wi-Fi chips in off-the-shelf smartphones, we cannot perform comparative micro-benchmarks using ready to use applications. Therefore, we adapted parts of the Linux Internet Control Message Protocol (ICMP) implementation. ICMP is used by nodes in the network to send control messages like indicating success or failure when communicating with other nodes. Since the Linux implementation is deeply embedded into the operating system kernel, it is nearly impossible to extract the entire code related to ICMP. Instead, we adapted the ICMP code for handling ICMP *echo* packets, as commonly found in the "ping" utility. Using our ICMP adaptation, we can measure basic power consumption and latency executed in three different environments, i. e., user space, operating system kernel, and Wi-Fi chip.

**Experimental Setup** We placed two Nexus 5 smartphones about 30 cm apart from each other. The first Nexus 5 sent ICMP *echo-requests*, while the second one processed the received frames using two different versions of our ICMP program, both returning an *echo-reply* at the end of the execution. The first version is implemented in pure C where ReactiFi was not involved and the ICMP *echo-requests* are handled in the same way as in the Linux kernel, sending an ICMP *echo-reply* without further computations. The second version of our ICMP program, written in ReactiFi, contains single *filter* (dropping non-ICMP *echo-requests*), *map* (computing the ratio between Wi-Fi frame size and ICMP packet size), and *fold* (counting the received ICMP *echo-requests*) reactives, and returns a corresponding *echo-reply* after the respective reactives. Each of the reactives were used in separate tests. The pure C version of the program enables us to compare the power consumption and latency to the ReactiFi implementation. During the experiments, MAC Protocol Data Unit Aggregation (A-MPDU) and frame re-transmission were disabled in the Wi-Fi firmware. Thus, every packet was sent once in a separate Wi-Fi frame, allowing an evaluation for each packet. To evaluate latency, we measured the ICMP round trip time for each packet. To get high-resolution time measurements without interference, an external device was used to capture ICMP packets between the two Nexus 5 smartphones using Wireshark.[2] To measure power consumption without interference from the battery, we removed the battery from the Nexus 5 and the charge controller from the battery, soldered wires to the charge controller, and put the controller without the battery back into the Nexus 5. The measurements were performed using a Monsoon High Voltage Power Monitor with a sample rate of 5 kHz and a resolution of 286 µA. The voltage was set to 4.2 V, which corresponds to about 92 % battery capacity.

**Power Consumption** Figure 9 shows the power consumption of the micro-benchmark. Each subplot illustrates a different test case, the x-axis shows the number of requests per second, grouped by execution environment. The y-axes denotes the power

[2] https://www.wireshark.org, last accessed 2020-10-01.

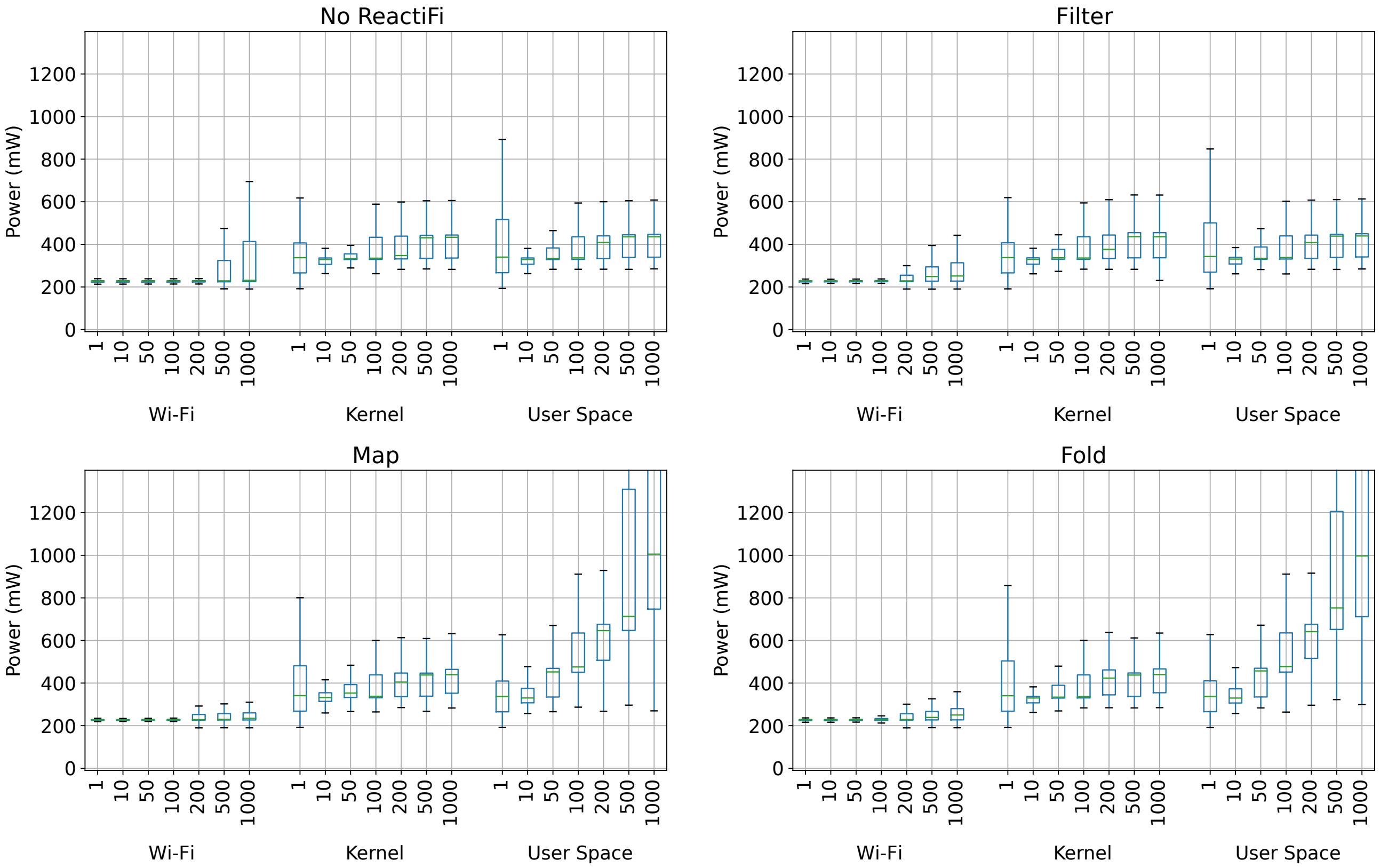

■ **Figure 9** Power consumption: Wi-Fi, OS kernel, and user space.

consumption. Note that for user space tests with high data rates, the boxes where truncated so that the Wi-Fi based tests can be seen better.

The ICMP program executed on the Wi-Fi chip saves up to 73 % power compared to the user space implementation and 30 % compared to the in-kernel execution, regardless of which program version was used, since the main CPU, which consumes more power in general, has to process all frames. Given that for low packet rates the kernel falls into power saving mode and user space applications are suspended and both need to be woken up quite often, tests with 1 request per second need more power than tests with 10 requests per second. This effect disappears with higher packet rates, since the ICMP program executed in the kernel or in user space is not suspended anymore. These tests show that executing code on the Wi-Fi chip reduces the power consumption compared to execution in the kernel or in user space. Additionally, when comparing the results of the four Wi-Fi experiments shown in figure 9, it is evident that reactives do not introduce significant power overhead compared to the purely C-based tests (shown in the upper left subfigure of figure 9).

**Latency** Figure 10 shows the round trip times (RTT) of ICMP *echo-requests* and the corresponding *echo-replies*. The RTTs, when processing directly on the Wi-Fi chip, are always low at about 0.3 ms, regardless of the number of requests/s. When incorporating reactives, RTTs are slower (about 1.4 ms to 1.8 ms), though consistent

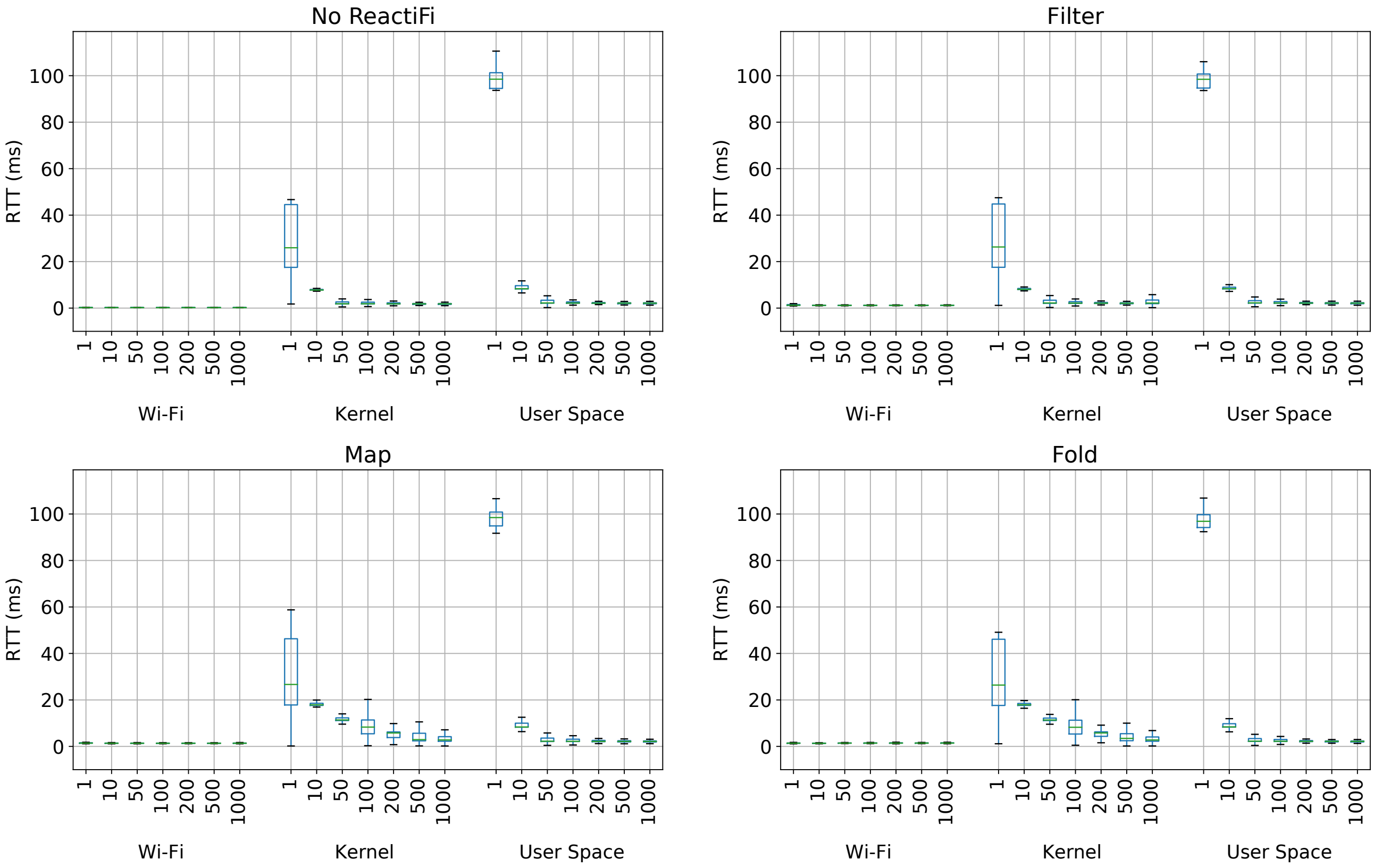

■ **Figure 10** ICMP Round Trip Time: Wi-Fi, OS kernel, and user space.

regardless of the data rates. The kernel and user space tests need 5 to 7 times more, depending on the test. With low data rates, however, both the kernel and the user space application are suspended, resulting in significantly higher RTTs (up to about 100 ms). These tests show that executing code on the Wi-Fi chip reduces the execution time, and thus, the overall latency vs. execution in the kernel or in user space. We can further observe that the execution time using the Wi-Fi chip is more predictable across the tests. Finally, compared to the purely C-based tests (shown in the upper left subfigure of figure 10), ReactiFi does not introduce significant latency overhead.

### 6.2 Power Consumption of Counting Nearby Devices: Wi-Fi Chip versus User Space

This experiment empirically validates our claims related to power consumption. We compare the power consumption of our nearby-device counting case study (cf. section 2) when run on the Wi-Fi chip versus being executed in user space.

**Experimental Setup** We used a Nexus 5 smartphone in a controlled environment to identify the number of devices around the phone. Five other devices (laptops and Rapsberry Pis) were distributed around the room, so that a constant and verifiable count could always be guaranteed. We implemented the described application on the Wi-Fi firmware using ReactiFi as discussed in section 2. We executed the code for

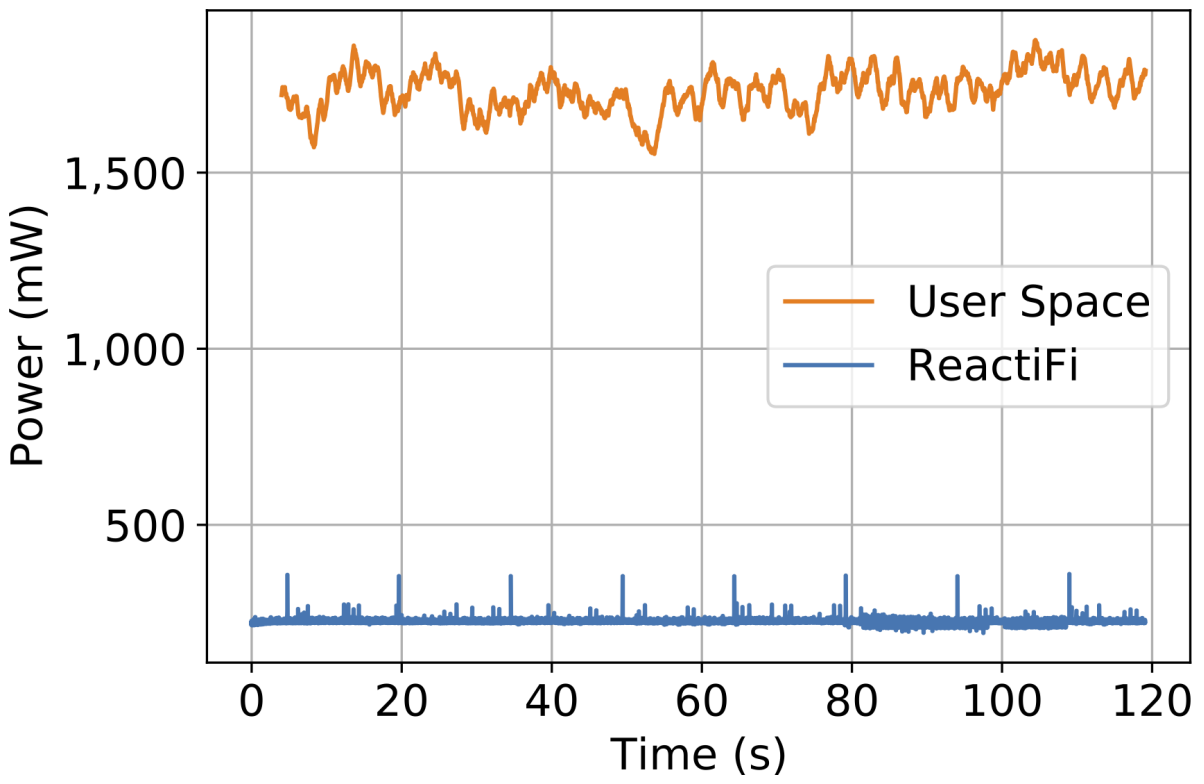

**(a)** Power consumption of the nearby-device counting implementations.

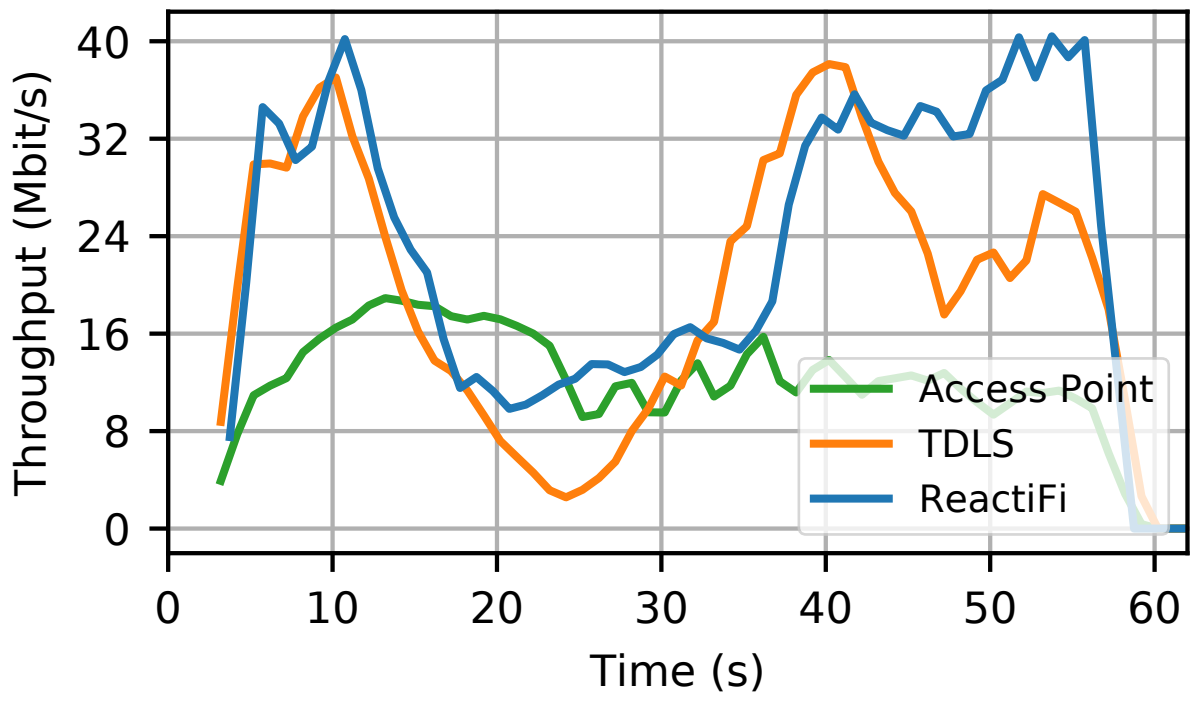

**(b)** Throughput of the adaptive file sharing application.

**Figure 11** Experimental results of the case studies.

120 s and repeated the experiment 5 times. Furthermore, we used the intermediate generated C code to execute the same case study in user space. Both implementations produced the same results in terms of the number of counted devices.

**Power Consumption** Figure 11a shows the average power consumption for both implementations over all runs. The code executed in user space requires about 1700 mW power on the average with 630 mW standard deviation. The ReactiFi program uses only 225 mW, thus, achieving 87 % improvement compared to the user space implementation. The peaks every 15 s are due to the LTE interface trying to connect to a network even if no SIM card is present.

### 6.3 Throughput Boosting by Adaptive File Sharing

This experiment empirically validates our claim that making the Wi-Fi chip programmable enables novel networking applications with improved throughput, such as our adaptive file sharing case study.

**Experimental Setup** We used the ReactiFi program in listing 2 on a Nexus 5 smartphone (receiver) to download a file from a Raspberry Pi 3 (sender) in the scenario shown in figure 1. The file is served by a standard HTTP server without modifications. We used a Turris Omnia RTROM01 router in stock configuration as our AP. Wi-Fi was set to IEEE 802.11n mode on channel 6 in the 2.4 GHz band to increase the usable Wi-Fi range.

As illustrated in figure 1, the AP was about 6 meters away from the stationary Nexus 5. In the beginning, the Raspberry Pi 3 was about 1 meter away from the Nexus 5 at $t = 0$, we then moved up to 3 m towards the AP ($t = 18$) and continued farther away from both the AP and the Nexus 5. The maximum distance between Nexus 5 and Raspberry Pi 3 was about 12 meters and 7 meters between Raspberry Pi 3 and AP ($t = 25$). After that, the same path was used for the way back ($t = 35$), resulting in the same position as at the beginning of the test ($t = 55$). The experiment took 55 s in total, while the maximum distance between Nexus 5 and Raspberry Pi 3 was reached after about 25 s. Since all surrounding wireless traffic had to be analyzed for this application, the Nexus 5 was set to Wi-Fi monitor mode.

**Throughput** Figure 11b shows throughput in Mbit/s (y-axis) and time in seconds (x-axis) during three tests: (i) using only the AP shown in figure 1 in IEEE 802.11n AP mode, (ii) using only IEEE 802.11z TDLS to establish a direct connection between the Nexus 5 and the Raspberry PI 3, and (iii) using the ReactiFi adaptive file sharing application to automatically switch between (i) and (ii).

While the throughput of test (i) in AP mode is more or less at about 12 Mbit/s during the entire test, the same file shared via TDLS in test (ii) shows peaks at about 40 Mbit/s at the beginning and the end of the test, i. e., when the Raspberry Pi 3 is close to the Nexus 5 ($t = 0$ and $t = 55$). At the maximum distance (i. e., the worst SNR) between the devices ($t = 25$), the throughput drops to 4 Mbit/s in the TDLS-only test.

The same experiment performed with the ReactiFi program in test (iii) results in significant improvements over both the AP and TDLS tests, as shown in figure 11b. At the beginning and at the end (i. e., with the best SNR), the results are comparable to the TDLS-only test (ii), where throughput exceeds 40 Mbit/s. However, with the worst SNR, the throughput does not fall below the values of the AP-only test (i). This experiment shows that ReactiFi enables the development of novel Wi-Fi applications that cannot be implemented in the operating system kernel or in user space and that can have significant performance gains, leading to up to a factor of 3.3 higher throughput compared to AP-only mode.

## 7 Related Work

**Extending Network Functionality** Using eBPF to program the Linux kernel network stack, multiple works propose extension systems for IPv6 [27, 56], OSPF [55], TCP [8, 27, 49], Multipath TCP [22], BGP [55], and QUIC [16]. All these approaches, however, rely on the C language with its downsides. Additionally, these approaches cannot access information from the PHY and MAC layers.

**Programmable Wi-Fi Firmware** Tinnirello, Bianchi, Gallo, Garlisi, Giuliano, and Gringoli [48] present a finite state machine approach for defining and implementing MAC protocols on Wi-Fi firmware, and Bianchi, Gallo, Garlisi, Giuliano, Gringoli, and Tinnirello [6] extend this approach by introducing MAClets for simplifying the programmability of MAC protocols executed on Wi-Fi firmware. In contrast, ReactiFi supports a wide range of applications not limited to MAC protocols. Furthermore, ReactiFi runs on off-the-shelf wireless devices, whereas MAClets require the software-defined radio platform USRP B200 for execution.

**Software-defined Wireless Networking** Software-defined networking (SDN) [9, 10, 20, 24] supports programmable network behavior in a centrally controlled manner to facilitate flexible network management. Several languages and systems can be used to program the data plane of SDN switches, such as P4 [7] or OpenFlow [31], and wireless networks [4, 15]. Programmability of wireless networks is promising especially at the PHY and MAC layers due to the dynamicity of wireless communication and the scarcity of the wireless spectrum [19, 28]. Schulz-Zander, Mayer, Ciobotaru, Schmid, and Feldmann have proposed OpenSDWN, an approach based on SDWN and Network Function Virtualization (NFV) [44, 45, 46]. In OpenSDWN, a virtual AP is provided for each client, where PHY and MAC layer transmission settings can be changed for each flow. Hätönen, Savolainen, Rao, Flinck, and Tarkoma [25] use intelligent edge techniques to enable SDWN on off-the-shelf APs, where virtual machines are used on AP hardware to create multiple virtual APs.

To the best of our knowledge, no existing SDWN approach facilitates the programmability of Wi-Fi functionality on off-the-shelf mobile devices, as it is supported by ReactiFi. Furthermore, we argue that SDWN approaches are not well-suited for the goal of programming Wi-Fi firmwares of mobile devices, because programming wireless networks is about reacting to events (e. g., timers, incoming frames, OS control signals), and maintaining state about a device's physical context (e. g., its neighbors and their distances), not only about packet processing.

**Event-based Embedded Programming** TinyOS [29] is a scheduler and a collection of drivers for low-power wireless embedded systems. It allows event-driven programming with nesC [23], a C language derivative. RIOT OS [2] is a microkernel-based operating system, designed to match the requirements of Internet of Things (IoT) devices. It allows thread execution with a preemptive, priority-based scheduler, but does not include integrated means to handle dataflow.

**Reactive Programming for Embedded Systems** Emfrp [39, 53], CFRP [47], and Hae [52] are reactive programming languages for generic embedded devices, often modeling sensor-based devices that monitor some external state. This leads to a design that focuses only on stateful reactives, i. e., the value of a sensor can always be accessed. Thus, these approaches lack built-in operations such as filtering and alternatives of events. They have been shown to be easily parallelizable [34].

Juniper [26] is an ML-like language for the Arduino platform. While it supports both stateful and stateless events, Juniper does not distinguish the two, blurring the semantics with regards to when an event fires and where the runtime has to store state. Juniper supports a dynamic dataflow graph by compiling a runtime into the target C++ code, resulting in a more complex program with a larger memory footprint. It also allows inline C++ code, similar to how ReactiFi is based on C, but both C++ and Juniper have redundancies, and the interaction between the Juniper code and the C++ code require understanding of the encodings by the Juniper compiler.

Flask [30] uses the Haskell type system to type an embedded DSL similar to ReactiFi, showing the applicability of the approach to other domains. However, since Flask targets sensor networks, the semantics of Flask are optimized for a system that allows less control, leading to a language with fewer guarantees.

CÉU [38] and Esterel [5, 18] do not directly focus on embedded devices, but bring synchronous reactive programming to soft real-time systems. Compared to ReactiFi they use a more imperative style of syntax, but their underlying semantics, i. e., processing all events in the same logical time step, is similar to ReactiFi. These languages are used in industry deployments, demonstrating the advantages over alternatives such as writing C directly.

**Reactive Programming for Programmable Networks** Frenetic [21], Nettle [50] and Procera [51] allow programmers to describe network policies using functional reactive programming abstractions. Compared to ReactiFi, these languages target packet forwarding on programmable network switches, by compiling to OpenFlow [31] rules. The dataflow in these languages is very specific to the semantics of OpenFlow and does not directly translate to Wi-Fi programming. Flowlog [33] adopts a database-like programming model, where internal state, represented as tables, is updated in response to incoming events. The SQL-like syntax of Flowlog hides the dataflow of applications and makes it hard to compose independent event flows without creating intermediate tables.

## 8 Conclusion

We presented ReactiFi, a domain-specific language to facilitate programmability of Wi-Fi firmware on mobile consumer devices. Using ReactiFi, programmers use a high-level reactive programming language to operate on PHY, MAC and IP layer mechanisms, such as reception of frames or changed radio link properties. Developers implement applications that access information available only in the Wi-Fi firmware in a high-level language that compiles to efficient binary code. We discussed the advantages of ReactiFi with respect to scheduling, memory usage, and basic Wi-Fi functionality, and by comparing two implementations of a case study in C and ReactiFi. Our empirical evaluation demonstrated the benefits of programming Wi-Fi firmware in terms of significant improvements of throughput, latency, and power consumption.

There are several areas for future work. For example, it would be interesting to evaluate ReactiFi in other usage scenarios based on 60 GHz Wi-Fi networks, or transferring it to other wireless technologies, such as WiMAX, 4G, LoRa, Bluetooth, or ZigBee. Furthermore, ReactiFi should be integrated into SDWN frameworks, which would allow new types of control and novel applications in a wide range of scenarios. Supporting an execution environment like eBPF could help to create secure platforms, where unknown code could be executed with limited security implications, allowing platforms for distributing and sharing arbitrary ReactiFi firmware programs. On the language side, the ReactiFi DSL could serve as a bridge between user space and Wi-Fi programs, allowing to program parts of an application on a Wi-Fi chip and other parts in user space using a single code base and removing boilerplate code. Another interesting area is to extend the type system with assume-guarantee reasoning, by letting the programmer provide high-level specifications of the memory and real-time characteristics of the user-defined functions. This could then be used to harden the guarantees of the entire application. Finally, user studies should be conducted to further strengthen the arguments about the benefits of ReactiFi.

**Acknowledgements** This work is funded by the German Research Foundation (Project 415626024 and SFB 1053), by the European Research Council (Advanced Grant 321217 and 862535), by the German Federal Ministry of Education and Research together with the Hessen State Ministry for Higher Education (ATHENE), and LOEWE in Hessen, Germany (emergenCITY, Natur 4.0).

## Appendix

**Table 1** Predefined interactions between ReactiFi application and runtime.

| Category | Declaration | Description |
|---|---|---|
| Receive | SentFrame | Outgoing frame |
| | ReceivedFrame | Incoming frame filtered by destination addr. |
| | Monitor | All incoming frames |
| Management | ScanResult | Results of a Wi-Fi access point scan |
| | ChannelState | Updated channel state information |
| | TxPower | Updated transmit power information |
| Interrupts | IOCTL | ioctl data |
| | Timer | Timer event with ms granularity |
| Effects | SendFrame | Send Wi-Fi frame |
| | SwitchChannel | Switch Wi-Fi channel |
| | ChangeCSI | Change Wi-Fi channel state information |
| | SetTxPower | Set transmit power |
| | SendToOS | Send a value frame to the operating system |
| | SetTDLS | Change the connection mode to/from TDLS |

**Listing 6** C code of the adaptive file sharing case study.

```
#define FROM_AP 0
#define TO_AP 1
#define FROM_TDLS 2

#define FC_OFFSET 6
#define DURATION_OFFSET 8
#define ADDR1_OFFSET 10
#define ADDR2_OFFSET 16
#define ADDR3_OFFSET 22
#define SEQ_CTL_OFFSET 28

#define IOV_SET 1

typedef struct {
    uint8_t addr[6];
} mac_addr_t;

typedef struct {
    uint8_t version;
    uint8_t type;
    uint8_t sub_type;
    uint8_t to_ds;
    uint8_t from_ds;
    uint8_t more_frags;
    uint8_t retry;
    uint8_t pwr_mngmt;
    uint8_t more_data;
    uint8_t protected;
```

```
    uint8_t order;
} frame_control_t;

typedef struct {
    frame_control_t *fc;
    uint16_t duration;
    mac_addr_t *src;
    mac_addr_t *dst;
    mac_addr_t *bssid;
    uint16_t seq_ctl;
    int32_t signal;
    int32_t noise;
    uint8_t ds_type;
    struct wl_info *wl;
} monitor_frame_t;

struct averages {
    int32_t avg1;
    int32_t avg2;
    mac_addr_t *addr;
    struct wl_info *wl;
};

uint8_t is_tdls = 0;
map_t addr_snr_map = 0;
int *signal_count = 0;
uint8_t MY_MAC[6] = {0};

// If we receive a sk_buff, we have to parse it.
// This is what this function is for.
int make_frame(monitor_frame_t *mntr, struct wl_info *wl, struct wl_rxsts *sts, struct sk_buff *p)
    ↪ {

    char *raw_frame = (char *)p->data;

    mntr->fc->version = (uint8_t) (raw_frame[FC_OFFSET] & 0x03);
    mntr->fc->type = (uint8_t) (raw_frame[FC_OFFSET] & 0x0C) >> 2;
    mntr->fc->sub_type = (uint8_t) (raw_frame[FC_OFFSET] & 0xF0) >> 4;

    if (mntr->fc->type != 2) {
        return -1;
    }

    if (mntr->fc->sub_type != 0) {
        return -1;
    }

    mntr->fc->to_ds = (uint8_t) (raw_frame[FC_OFFSET + 1] & 0x01);
    mntr->fc->from_ds = (uint8_t) (raw_frame[FC_OFFSET + 1] & 0x02) >> 1;
    mntr->fc->more_frags = (uint8_t) (raw_frame[FC_OFFSET + 1] & 0x02) >> 2;
    mntr->fc->retry = (uint8_t) (raw_frame[FC_OFFSET + 1] & 0x02) >> 3;
    mntr->fc->pwr_mngmt = (uint8_t) (raw_frame[FC_OFFSET + 1] & 0x02) >> 4;
    mntr->fc->more_data = (uint8_t) (raw_frame[FC_OFFSET + 1] & 0x02) >> 5;
```

```
    mntr->fc->protected = (uint8_t) (raw_frame[FC_OFFSET + 1] & 0x02) >> 6;
    mntr->fc->order = (uint8_t) (raw_frame[FC_OFFSET + 1] & 0x02) >> 7;

    memcpy(&mntr->duration, &raw_frame[DURATION_OFFSET], 2);

    if (mntr->fc->to_ds == 0 && mntr->fc->from_ds == 1) {
        mntr->ds_type = FROM_AP;
        memcpy(mntr->dst, &raw_frame[ADDR1_OFFSET], 6);
        memcpy(mntr->dst, &raw_frame[ADDR3_OFFSET], 6);
    }
    // REPLY TO AP
    else if (mntr->fc->to_ds == 1 && mntr->fc->from_ds == 0) {
        mntr->ds_type = TO_AP;
        memcpy(mntr->dst, &raw_frame[ADDR3_OFFSET], 6);
        memcpy(mntr->dst, &raw_frame[ADDR2_OFFSET], 6);
    }

    // REPLY TDLS
    else if (mntr->fc->to_ds == 0 && mntr->fc->from_ds == 0) {
        mntr->ds_type = FROM_TDLS;
        memcpy(mntr->dst, &raw_frame[ADDR1_OFFSET], 6);
        memcpy(mntr->dst, &raw_frame[ADDR2_OFFSET], 6);
    }

    else {
        return -1;
    }

    memcpy(&mntr->seq_ctl, &raw_frame[SEQ_CTL_OFFSET], 2);

    mntr->signal = sts->signal;
    mntr->noise = sts->noise;

    mntr->wl = wl;

    return 0;
}

// Function for generating the key for the hashmap
void gen_key(char *key, uint8_t type, monitor_frame_t *input) {
    sprintf(
        key, // Store in key
        "%d %x%x%x%x%x%x", // Format: type, space, 6 byte mac address
        type,
        input->src->addr[0],
        input->src->addr[1],
        input->src->addr[2],
        input->src->addr[3],
        input->src->addr[4],
        input->src->addr[5]
    );
}

```

```

// The first filter for filtering frames sent to this device.
monitor_frame_t *filter_my_frames(monitor_frame_t *input, uint8_t MY_MAC[6]) {
    // Compare the 6 bytes containing the dst address with my_mac_address
    // If it is the same, yield the next reactive.
    // Otherwise, do nothing.
    if (memcmp(input->dst, MY_MAC, 6) == 0) {
        return input;
    } else {
        return 0;
    }
}

// Aggregates SIGNAL_COUNT signals for the current address to an average
map_t aggregate_average(map_t state, monitor_frame_t *input, int *signal_count) {
    char *key = malloc(15, 0);
    gen_key(key, input->ds_type, input);

    // Get the signals for the current address.
    int *avg = 0;
    int hashmap_state = hashmap_get(state, key, (void **)&avg);

    // If the address is not present, we have not seen this frame yet.
    // Initialize.
    if (hashmap_state != MAP_OK) {
        avg = malloc(sizeof(int), 0);
        *avg = 0;
    }

    // Calculate the average and put the new average to the map.
    *avg = *avg + (input->signal - *avg) / *signal_count;
    *signal_count = *signal_count + 1;

    // Put the newly created avg to the hashmap.
    hashmap_put(state, key, avg);

    return state;
}

// Comparing signals for deciding if TDLS should be setup/destroyed.
struct averages *compare_signals(map_t state, monitor_frame_t *input) {
    char *key = malloc(15, 0);
    gen_key(key, input->ds_type, input);

    int *avg = 0;
    int snr_avgs_state = hashmap_get(state, key, (void **)&avg);

    // If the key is not in the hashmap, something is wrong. Abort.
    if (snr_avgs_state != MAP_OK) {
        return 0;
    }

    char *other_key = malloc(15, 0);
```

```
    int *other_avg = 0;
    switch (input->ds_type) {
        case FROM_AP:
        case TO_AP:
            // If this frame if from the AP or to the AP, we compare it to the SNR average received
                ↪ directly from the other node.
            // If it is smaller, enable TDLS.
            gen_key(other_key, FROM_TDLS, input);

            snr_avgs_state = hashmap_get(state, other_key, (void **)&other_avg);
            if (snr_avgs_state != MAP_OK) {
                return 0;
            }

            break;

        case FROM_TDLS:
            // If this frame is directly from the other node, we compare it to the SNR average
                ↪ received from the AP.
            // If it is bigger, enable TDLS.
            gen_key(other_key, FROM_AP, input);

            snr_avgs_state = hashmap_get(state, other_key, (void **)&other_avg);
            if (snr_avgs_state != MAP_OK) {
                return 0;
            }

            break;

        default:
            return 0;
    }

    free(key);
    free(other_key);

    // Construct the return struct
    struct averages *ret = (struct averages*) malloc(14 + sizeof(struct wl_info*), 0);
    ret->avg1 = *avg;
    ret->avg2 = *other_avg;
    ret->addr = input->src;
    ret->wl = input->wl;

    return ret;
}

// Filter for enabling TDLS if it should be.
void enable_tdls(struct averages *avgs) {
    if (avgs->avg1 < avgs->avg2) {
        // Before enabling TDLS, check if it already is enabled.
        if (is_tdls == 0) {
            struct tdls_iovar info;
            memset(&info, 0, sizeof(struct tdls_iovar));
```

```
            memcpy(info.ea, avgs->addr, 6);

            info.mode = TDLS_MANUAL_EP_DISCOVERY;
            wlc_iovar_op(avgs->wl->wlc, "tdls_endpoint", 0, 0, &info, sizeof(struct tdls_iovar), IOV_SET, 0);

            info.mode = TDLS_MANUAL_EP_CREATE;
            wlc_iovar_op(avgs->wl->wlc, "tdls_endpoint", 0, 0, &info, sizeof(struct tdls_iovar), IOV_SET, 0);

            is_tdls = 1;
        }
    }
}

// Filter for disabling TDLS if it should be.
void disable_tdls(struct averages *avgs) {
    if (avgs->avg1 > avgs->avg2) {
        // Before disabling TDLS, check if is enabled.
        if (is_tdls == 1) {
            struct tdls_iovar info;
            memset(&info, 0, sizeof(struct tdls_iovar));
            memcpy(info.ea, avgs->addr, 6);

            info.mode = TDLS_MANUAL_EP_DELETE;
            wlc_iovar_op(avgs->wl->wlc, "tdls_endpoint", 0, 0, &info, sizeof(struct tdls_iovar), IOV_SET, 0);

            is_tdls = 1;
        }
    }
}

// The monitor source function.
// Whenever a new frame appears, we do the handling here.
void wl_monitor_hook(struct wl_info *wl, struct wl_rxsts *sts, struct sk_buff *p) {
    if (p == 0 || p->data == 0) {
        return;
    }

    monitor_frame_t *frm = (monitor_frame_t *) malloc(42 + sizeof(struct wl_info*), 0);
    frm->fc = (frame_control_t*) malloc(11, 0);
    frm->dst = (mac_addr_t*) malloc(6, 0);
    frm->src = (mac_addr_t*) malloc(6, 0);
    frm->bssid = (mac_addr_t*) malloc(6, 0);

    if (make_frame(frm, wl, sts, p) != 0) {
      goto cleanup;
    }

    monitor_frame_t *my_frames = filter_my_frames(frm, MY_MAC);
    if (!my_frames) {
      goto cleanup;
```

```
	}

	addr_snr_map = aggregate_average(addr_snr_map, my_frames, signal_count);

	struct averages *tmp3 = compare_signals(addr_snr_map, frm);
	if (!tmp3) {
		goto cleanup;
	}

	enable_tdls(tmp3);
	disable_tdls(tmp3);

cleanup:
	free(frm->fc);
	free(frm->dst);
	free(frm->src);
	free(frm->bssid);
	free(frm);
	wl_monitor(wl, sts, p);
}

// This is the firmware's main function.
// Initialize the hashmap and the counter.
// Enable monitor mode as the frame source.
void autostart(int a1) {
	addr_snr_map = hashmap_new();
	signal_count = malloc(sizeof(int), 0);
	*signal_count = 1;

	__attribute__((at(0x18DA30, "", CHIP_VER_BCM4339, FW_VER_6_37_32_RC23_34_40_r581243)))
	__attribute__((at(0x18DB20, "", CHIP_VER_BCM4339, FW_VER_6_37_32_RC23_34_43_r639704)))
	BLPatch(wl_monitor_hook, wl_monitor_hook);
}
```

## About the authors

**Artur Sterz** is a PhD student in the Department of Mathematics and Computer Science at Philipps-Universität Marburg, Germany. His current research focuses on wireless communication in mobile networks. Contact him at sterz@informatik.uni-marburg.de

**Matthias Eichholz** is a PhD student in the Department of Computer Science at Technische Universität Darmstadt, Germany. His current research focuses on programming languages for software-defined networks. Contact him at eichholz@cs.tu-darmstadt.de

**Ragnar Mogk** is a PhD student in the Department of Computer Science at Technische Universität Darmstadt, Germany. His current research focuses on programming languages for distributed applications. Contact him at mogk@cs.tu-darmstadt.de

**Lars Baumgärtner** is a postdoctoral researcher in the Department of Computer Science at Technische Universität Darmstadt, Germany. His current research focuses on delay-tolerant networking. Contact him at baumgaertner@cs.tu-darmstadt.de

**Pablo Graubner** was a postdoctoral researcher in the Department of Mathematics and Computer Science at Philipps-Universität Marburg, Germany. His research focuses on energy-efficient computation. Contact him at graubner@informatik.uni-marburg.de

**Matthias Hollick** is a full professor in the Department of Computer Science at Technische Universität Darmstadt, Germany. His current research interests are wireless communication and secure mobile networking. Contact him at mhollick@seemoo.tu-darmstadt.de

**Mira Mezini** is a full professor in the Department of Computer Science at Technische Universität Darmstadt, Germany. Her current research interests are programming languages and software engineering. Contact her at mezini@st.informatik.tu-darmstadt.de

**Bernd Freisleben** is a full professor in the Department of Mathematics and Computer Science at Philipps-Universität Marburg, Germany. His current research interests are distributed systems, mobile computing, and networked applications. Contact him at freisleb@informatik.uni-marburg.de